# TAC PROPOSAL FOR FUNDAMENTAL AND APPLIED RESEARCH [1]

A.K. Çiftçi[1], S. Sultansoy[2,3,5], Ö. Yavaş[4], M. Yılmaz[5]

[1] Ankara Üniversitesi, Fen Fak., Fizik Böl., 06100 Tandoğan, Ankara, Türkiye
[2] Deutsches Elektronen-Synchrotron, Notke Str. 85, 22607 Hamburg, Germany
[3] Elmler Akademiyası Fizika Institutu, H. Cavid prospekti 33, Bakı, Azerbaycan
[4] Ankara Üniversitesi, Fen Fak., Fizik Müh. Böl., 06100 Tandoğan, Ankara, Türkiye
[5] Gazi Üniversitesi, Fen-Edeb. Fak., Fizik Böl., 06500 Teknikokullar, Ankara, Türkiye

**Abstract**

Main parameters of linac-ring type $\phi$ factory and 1÷5 GeV energy proton synchrotron proposed in the framework of the Turkic Accelerator Complex (TAC) Project are discussed.

**Linac-ring type $\phi$ factory.** Two sets of parameters, corresponding to E=130 (260) MeV for electron linac and E=1000 (2000) MeV for positron ring, are considered. It is shown that $L=10^{34} cm^{-2} s^{-1}$ can be achieved, which is much higher than the design luminosity of the DA$\phi$NE (Frascati, Italy). Therefore, a number of fundamental problems of particle physics such as CP violation, rare decays of K-mesons *et cetera* can be investigated with highest statistics. Moreover, asymmetry of kinematics can be advantageous for measuring neutral K-meson's oscillations and CP violation parameters. Parameters of the FEL (Free Electron Laser) based on electron linac and synchrotron radiation sources based on the positron ring are estimated. These beamlines will give opportunity for a wide spectrum of applied research, which include molecular biology, material science, medicine *et cetera*.

**Proton synchrotron.** This part consists of 100÷300 MeV energy linear pre-accelerator and 1÷5 GeV main ring. Proton beams from two different points of the synchrotron will be forwarded to neutron and muon regions, where a wide spectrum of applied research is planned. In muon region, together with fundamental investigations such as test of QED and muonium-antimuonium oscillations, a lot of applied investigations such as High-$T_c$ superconductivity, phase transitions, impurities in semiconductors *et cetera* will be performed using the powerful Muon Spin Resonance ($\mu$SR) method. In neutron region investigations in different fields of applied physics, engineering, molecular biology and fundamental physics are planned. In addition, some principal aspects of recently proposed new type nuclear reactors with Thorium fuel could be studied.

---

[1] Talk given at the I. Eurasia Conference on Nuclear Science and its Applications, Izmir, Turkey, 23-27 October 2000

# 1. Introduction

Today, Accelerator Centers are cornerstones of the National R&D structure of developed countries. An impressive examples are KEK in Japan and Pohang AC in Korea. Unfortunately, Turkey has not such a Center. Moreover, concerning High Energy Physics (HEP) in general, in order to satisfy the European Union standards, Turkey should have ~250 theorists, ~400 experimental physicists and ~200 accelerator scientists with Ph.D. degree. However, actual numbers are ~50, ~10 and ~0, correspondingly. The analysis of the present state of the HEP in Turkey and new Turkic states, as well as the list of suggestions in order to promote of this field of science in our region can be found in [1].

The aim of the Turkic Accelerator Center (TAC) [2], which was proposed in the framework of the ATAM Science City Project and is supported by the Turkish State Planning Organisation under the Grant No DPT-97K-120420, is the filling of the above-mentioned "vacuum" in the most effective (and cheapest) manner. For this reason, we have consider the most compact accelerator complex, which will give an opportunity to perform both fundamental and applied researches. Below, we present recent state of our work on this subject.

# 2. Linac-ring type $\phi$-factory

The old idea [3] to collide a beam from a linear accelerator with a beam circulating in a storage ring has been renewed recently for two purposes: to achieve the TeV scale in lepton-hadron and photon-hadron collisions (see review [4] and references therein) and to construct high luminosity particle factories. In the last direction linac-ring type B-factory [5], c-$\tau$-factory [6] and $\phi$-factory [7] have been proposed. Here we present some preliminary results on the linac-ring type $\phi$-factory studies performed by the Ankara University Accelerator Physics Group [2, 8 and 9].

A general scheme of the proposed complex is given in Figure 1. Electrons accelerated in main linac up to energies 260(130) MeV are forwarded to detector region where they collide with positrons from main ring or turned out to undulator region where FEL beam is produced. On the other side electrons, accelerated in a small linac, are forwarded to conversion region where positron beam is produced. Then, positrons are accumulated in booster and after some beam gymnastics are forwarded to the main ring and accelerated up to energies 1(2) GeV. Wigglers installed in two regions will provide SR for applied researches.

Main parameters of proposed machine are given in Table 1 for two different choices of electron and positron beam energies. Below we present several illuminating notes. Electron bunches accelerated in main linac are used only once for collisions. On the other hand, positron bunches have to be used numerously, therefore, the stability of positron beam is very important. Empirically, allowed values for beam-beam tune shift is $\Delta Q \leq 0.06$ for lepton beams in storage rings. In principle, this upper limit taken from experiments done in usual (ring-ring type) $e^+e^-$ colliders can be higher for linac-ring type machines. Nevertheless, we use the conservative value $\Delta Q \leq 0.06$ for both options. The smallness of fractional energy loss of electrons in positron beam field is very important for resonance production of $\phi$-mesons. For this reason $\delta$ should be less than 0.004. The usage of the flat beams may be advantageous in order to reduce the huge value of disruption parameter for the electrons. The work on the subject is under development.

Table I. Main parameters of the φ factory

| Electron beam energy, MeV | 130 | 260 |
|---|---|---|
| Positron beam energy, MeV | 2000 | 1000 |
| Center of mass energy, MeV | 1020 | 1020 |
| Radius of ring, m | 50 | 30 |
| Acceleration gradient, MV/m | 12.5 | 12.5 |
| Length of main linac, m | 10 | 20 |
| Electrons per bunch, $10^{10}$ | 0.04 | 0.02 |
| Positrons per bunch, $10^{10}$ | 10 | 20 |
| Collision frequency f, MHz | 30 | 30 |
| Bunches per ring | 32 | 19 |
| Electron current, mA | 1.92 | 0.96 |
| Positron current, A | 0.96 | 0.48 |
| Energy loss per turn, keV | 30 | 3 |
| Fractional energy loss of the electrons δ, $10^{-4}$ | 2 | 1 |
| Beam size at the collision point $\sigma_{x,y}$, μm | 1 | 1 |
| Luminosity L, $10^{34}$cm$^{-2}$s$^{-1}$ | 1 | 1 |

Since deviation of the center-of-mass energy of $e^+e^-$ collisions is smaller than the total decay width of φ-meson, cross-section in the φ resonance region can be taken as σ≈4.4·$10^{-30}$cm$^2$. In the proposed complex 4.4·$10^{11}$ φ-mesons, 2.2·$10^{11}$ K$^+$K$^-$ pairs and 1.5·$10^{11}$ K$^0$K$^0$ pairs can be produced in a working year ($10^7$s). Fundamental problems of particle physics such as CP violation, rare decays of K-mesons etc. can be investigated with highest statistics. Moreover, kinematics asymmetry can be advantageous for measuring neutral K-meson's oscillations and CP violation parameters.

Table II. Event numbers in proposed collider

| Decay channels | Branching ratios | LR factory |
|---|---|---|
| K$^+$K$^-$ | 0.495 | 2.2·$10^{11}$ |
| K$^0$K$^0$ | 0.344 | 1.5·$10^{11}$ |
| ρπ | 0.155 | 6.9·$10^{10}$ |
| ηγ | 1.26·$10^{-2}$ | 5.6·$10^9$ |
| π$^0$γ | 1.31·$10^{-3}$ | 5.8·$10^8$ |
| e$^+$e$^-$ | 2.99·$10^{-4}$ | 1.3·$10^8$ |
| μ$^+$μ$^-$ | 2.5·$10^{-4}$ | 1.1·$10^8$ |
| ηe$^+$e$^-$ | 1.3·$10^{-4}$ | 5.8·$10^7$ |
| π$^+$π$^-$ | 8·$10^{-5}$ | 3.6·$10^7$ |
| η′(958)γ | 1.2·$10^{-4}$ | 5.4·$10^7$ |
| μ$^+$μ$^-$γ | 2.3·$10^{-5}$ | 1.0·$10^7$ |

## 3. Synchrotron Radiation Facility

By inserting wigglers on the straight parts of the main ring of φ-factory, one can produce synchrotron radiation for applied researches. If one use SiCo type magnet: $B_m$=3.33 Tesla, b=5.47 and c=1.8 [10], where $B_m$ is peak value of magnet's field, b and c are constants related to used permanent magnets. Main parameters of SR facility for two options are given in Table III. For applications see [2, 8].

Table III. Main parameters of SR facility

| Positron energy, MeV | 1000 | 2000 |
|---|---|---|
| Maximum magnetic field, T | 1.054 | 1.054 |
| Current, A | 0.976 | 0.488 |
| Period, cm | 3.2 | 13.2 |
| Gap $g$, mm | 30 | 30 |
| Total magnet length, m | 2.112 | 2.112 |
| Total radiated power, kW | 1.44 | 2.90 |
| Critical energy, keV | 0.700 | 2.804 |
| Wiggler parameter | 12.99 | 12.99 |
| Spectral flux, Phot/s·mrad·0.1%bandw | $4.96 \cdot 10^{14}$ | $5.00 \cdot 10^{14}$ |
| Spectral central brightness, Phot/s·mrad$^2$ ·0.1%bandw | $4.95 \cdot 10^{14}$ | $9.98 \cdot 10^{14}$ |

## 4. Free Electron Laser facility

Free Electron Laser (FEL) is a mechanism to convert some part of the kinetic energy of relativistic electron beam into tunable, highly bright and monochromatic coherent photon beam by using undulators inserted in linear accelerators or synchrotrons [11]. Main parameters of FEL facility for two options are given in Table IV. For using thereafter, magnetic field is estimated to be 1.48 kG with b=5.47, c=1.8, $\lambda_u$=33mm and $g$=25mm. With these values, strength parameter of undulator is K=0.456. Figure 2 shows the dependence of FEL flux on photon energy for $E_{e^-}$=260MeV option. Here peaks are placed at odd harmonics and maximum values of fluxes are $7.56 \cdot 10^{13}$, $1.08 \cdot 10^{13}$ and $9.45 \cdot 10^{11}$ for n=1, 3 and 5, respectively. Obtained averaged brightness values of photon beam are given in Table IV. For list of applications see [2, 8].

Table IV. Main parameters of FEL facility

| Electron energy, MeV | 130 | 260 |
|---|---|---|
| Photon energy, eV | 4.07 | 16.30 |
| Laser wavelength, A$^o$ | 3044 | 761 |
| Current, mA | 1.92 | 0.96 |
| Particle per bunch, $10^{10}$ | 0.04 | 0.02 |
| Repetition frequency, MHz | 30 | 30 |
| Flux, Phot/s·mrad·0.1%bandw | $3.78 \cdot 10^{13}$ | $7.56 \cdot 10^{13}$ |
| Averaged brightness, Phot/s·mrad$^2$ ·0.1%bandw | $2.91 \cdot 10^{11}$ | $5.81 \cdot 10^{11}$ |

# 5. Proton synchrotron

This part consists of 100÷300 MeV energy linear pre-accelerator and 1÷5 GeV main ring. Very preliminary lists of parameters are given in Tables V and VI.

Table V. Main parameters of linear pre-accelerator

| Options | E = 100 MeV | E = 200 MeV | E = 300 MeV |
|---|---|---|---|
| Repetition frequency, Hz | 25 | 25 | 25 |
| Average current, mA | 1 | 1 | 1 |
| Length, m | 10 | 20 | 30 |

Table VI. Main parameters of synchrotron

| Options | E = 1 GeV | E = 3 GeV | E = 5 GeV |
|---|---|---|---|
| Repetition frequency, Hz | 25 | 25 | 25 |
| Peak magnetic field, T | 1 | 3 | 5 |
| Circumference, m | 100 | 100 | 100 |

Proton beams from two different points of the synchrotron will be forwarded to neutron and muon regions, where a wide spectrum of applied research is planned. In muon region, together with fundamental investigations such as test of QED and muonium-antimuonium oscillations, a lot of applied investigations such as High-$T_c$ superconductivity, phase transitions, impurities in semiconductors *et cetera* will be performed using the powerful Muon Spin Resonance (µSR) method. In neutron region investigations in different fields of applied physics, engineering, molecular biology and fundamental physics are planned. In addition, some principal aspects of recently proposed new type nuclear reactors with Thorium fuel could be studied. The work on both the parameter list and applications is under progress.

**Acknowledgements.** This work is supported by Turkish State Planning Organisation under the Grant No DPT-97K-120420 and DESY. S. Sultansoy is grateful to Professor C. Yalcin for his kind invitation to participate the I. Eurasia Conference on Nuclear Science and Its Applications.

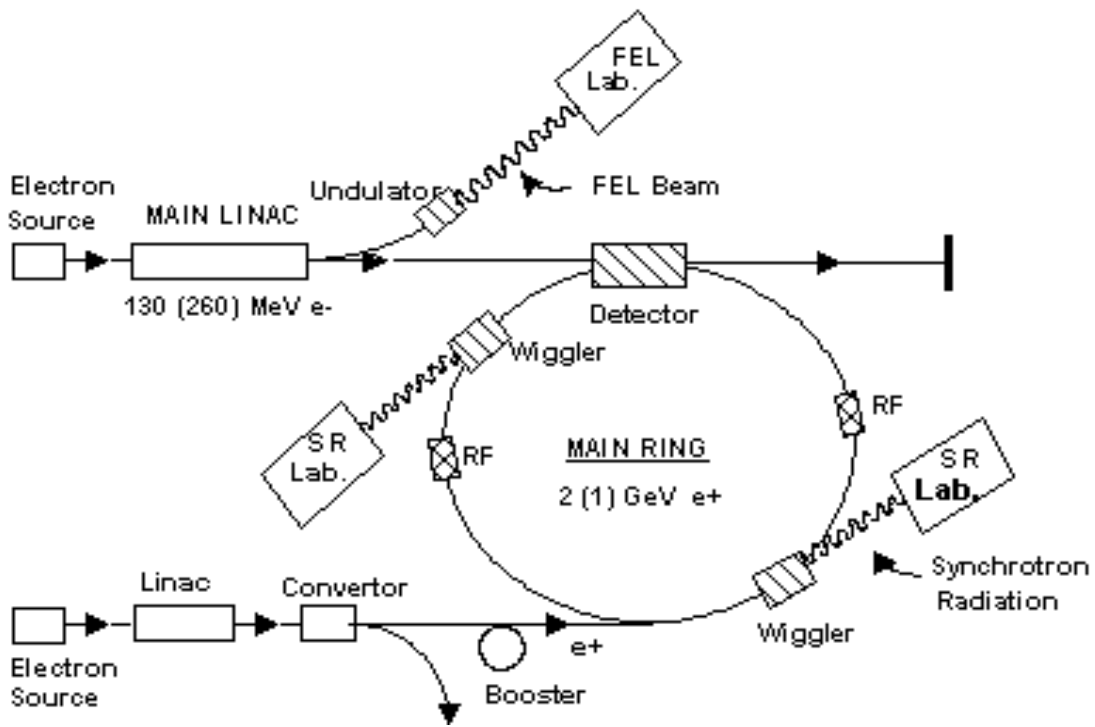

Figure 1. General scheme of the linac-ring type φ-factory

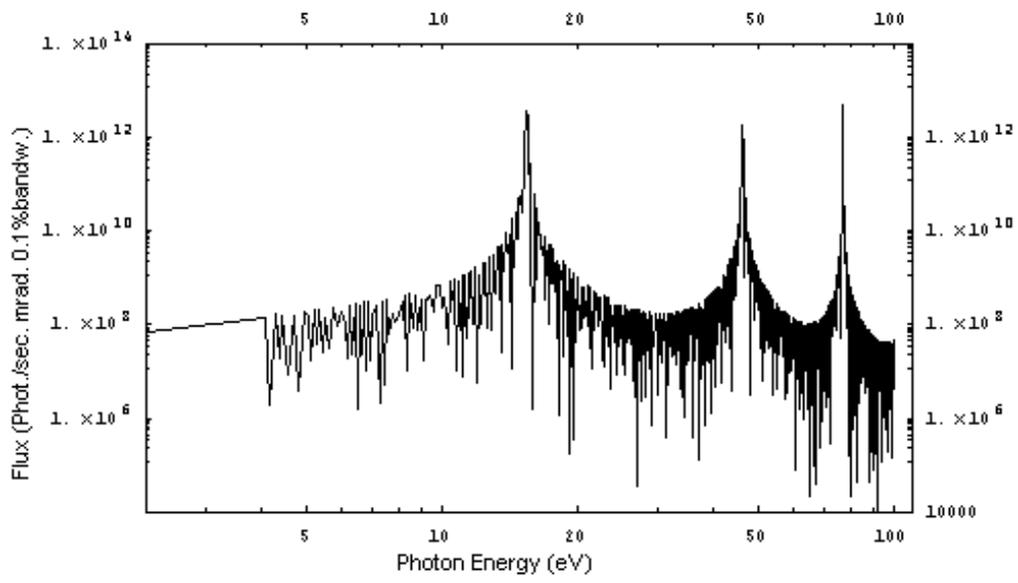

Figure 2. Flux of the FEL beam